\documentclass[a4paper,11pt]{article}
\usepackage{jinstpub} % for details on the use of the package, please see the JINST-author-manual
\usepackage{subcaption}
\usepackage{lineno}
\title{\boldmath A Neutron Sensitive Detector Using 3D-Printed Scintillators}

\author[a]{A. Barr,}
\author[a,b]{C. Da Vià,}
\author[a]{M.T. Binte Shawkat,}
\author[a]{M.J. Taylor,}
\author[a]{S. Watts,}
\author[a]{J. Allison,}
\author[c]{G. D'Amen,}
\author[d]{T. Gao}
\affiliation[a]{Department of Physics and Astronomy, University of Manchester,\\
Oxford Road, Manchester, United Kingdom}
\affiliation[b]{Department of Physics and Astronomy, Stony Brook University,\\
100 Nicolls Road, Stony Brook, United States of America}
\affiliation[c]{Brookhaven National Laboratory,\\
98 Rochester Street, Upton, United States of America}
\affiliation[d]{Department of Physics, University of Cambridge,\\
Trinity Lane, Cambridge, United Kingdom}

\emailAdd{adam.barr@manchester.ac.uk}

\abstract{This work reports on the performance of a novel neutron-sensitive scintillating detector fabricated using Fused-Deposition Modelling (FDM) additive manufacturing. FDM is a cost-effective 3D-printing method employing flexible plastic filaments to create custom-shaped components. Scintillating filaments, based on polystyrene doped with \emph{p}-terphenyl and 1,4-bis (5-phenyloxazol-2-yl) benzene,  and enriched with $^6$LiF to enable neutron sensitivity were manufactured in house and achieved visible scintillation with a light output of 30 $\pm$ 5~photons per MeV. Printed scintillators (both a 2~cm $\times$ 2~cm $\times$ 0.5~cm cuboid and a 2~cm $\times$ 2~cm $\times$ 1~cm cuboid containing a $^6$LiF-packed cavity) were then integrated into a detector system consisting of an image intensified TimePix3 camera, offering high spatial and temporal resolution. The detector performance was compared with Geant4 simulations of the scintillating sensor's response to electrons, gamma-rays, and thermal neutrons. A novel event discrimination algorithm, using the properties of the TimePix3 camera, enabled the separation of neutron signatures from the gamma-ray background.}


\arxivnumber{2507.08663} % Only if you have one

\begin{document}
\maketitle
\flushbottom

\section{Introduction}
\label{sec:intro}

Scintillators are a widely used solution for detecting radiation, offering high flexibility and a broad range of applications. The first plastic scintillator, polystyrene doped with \emph{m}-terphenyl, was discovered by Schorr and Torney in 1950~\cite{schorr_1950}. Since then, a large variety of scintillating materials have been discovered and used in particle physics, medicine, biology and other scientific fields in combination with visible light detectors~\cite{hamel2021}. 

The popularity of plastic scintillators is mainly due to their good light outputs combined with a rapid decay time making them highly suitable for particle identification. While many scintillators are composed of low-Z elements (mainly carbon and hydrogen), they could easily be combined with high-Z$_{\mathrm{eff}}$ materials to enable effective gamma-ray spectroscopy~\cite{Abreu_2018}. When layered with high-Z materials, scintillating materials have been used effectively for calorimetry~\cite{andragna_2006}. Segmented designs have also been used for particle tracking with submillimeter resolution~\cite{Masoliver2024}. 

Neutron-sensitive scintillators form a particular area of interest, with both scientific and commercial applications, for example in neutrino physics. Inverse beta decay, described by the following equation
\begin{equation}
\label{eq:idb}
\begin{aligned}
\bar\nu_e + \mathrm{p} \rightarrow \mathrm{n} + \mathrm{e}^+,
\end{aligned}
\end{equation}
sees both a neutron and a positron being produced~\cite{reines_cowan_1953}. In a scintillator with a short decay time, the time difference between the prompt scintillation light from the positron capture and delayed light from the neutron capture gives a clear signature of neutrino detection; this is a role to which plastic scintillators are well suited. As such, they have been used in a number of experiments, such as ISMRAN, VIDARR and the SuperFGD detector for the T2K experiment~\cite{dey2025,Bridges_2022, Kikawa2025}. Neutron-sensitive plastic scintillators have also found important applications in nuclear security and forensics. Neutrons are a key signature of fissile materials' decay, and their presence could pose major security threats if used for aggressive purposes; for this reason, there is extensive interest in neutron detection for homeland security and non proliferation (e.g.~\cite{gilliam_2002,bravar_2006, runkle_2010}). Furthermore, the resilience of plastic scintillators to changes in temperature and humidity allows their use in a wide variety of climatic conditions such as ports and countries' border crossings~\cite{hamel2021}. 

Maximizing the performance of plastic scintillators often demands complex geometries, which are difficult to achieve using traditional manufacturing techniques. In this article, we present a more versatile approach based on additive manufacturing for producing neutron-sensitive plastic scintillators. We also detail strategies to integrate these scintillators into a functional detector system employing a fast optical camera. This detector system is widely applicable, but is particularly targeted at nuclear security applications.

\section{Plastic Scintillators}
\label{sec:scintillators}

Most plastic scintillators exploit one of two main polymers: poly-vinyl toluene (PVT) or polystyrene (PS). These have very short attenuation lengths for optical photons, so must be combined with wavelength-shifting dyes. Several of these fluorophores are in common use, with \emph{p}-terphenyl (PTP), 2,5-diphenyloxazole (PPO), napthalene and 1,4-bis(5-phenyl-2-oxazolyl)benzene (POPOP) being commonly used~\cite{hamel2021}. While new plastics and fluorophores are continually being tested, few have been able to supplant these well-understood materials. 

Plastic scintillators have a well-understood mechanism of operation. The plastic can be understood as a matrix of polymers. Energetic charged particles entering the scintillator excite this matrix. The Foerster mechanism, a resonant dipole-dipole interaction, transfers this excitation to a primary fluorophore, such as PTP~\cite{birks_1960}. This nonradiative transfer depends heavily on the intramolecular distance. Once excited, the primary fluorophore emits photons. The attenuation length for scintillators with only a primary fluorophore is typically too low to be useful, less than 10~cm~\cite{hamel2021}. To avoid this, a secondary fluorophore, such as POPOP, is added. This absorbs photons emitted by the primary fluorophore and re-emits them at a more suitable wavelength~\cite{koshimizu_2025}. This also allows the emission wavelength of the scintillator to be tuned to meet the requirements of the photo-multiplier tube (PMT), silicon photo-multiplier (SiPM) or camera used to observe it. However, most secondary fluorophores emit in the blue band, so detectors should be sensitive to this band.

\subsection{Production of Scintillating Materials}
\label{sec:scint_prod}

The fabrication of plastic scintillators usually starts with bulk monomers, primarily styrene or vinyltoluene. These are heated to induce polymerization. If fluorophores are added to the monomers before heating, the resulting liquid polymer can be poured into moulds and cast to form a scintillator~\cite{hamel2021}. This produces a block of high-quality scintillator, but the equipment required is expensive and the process requires careful temperature control. 

Another option is to use granules of the base polymer combined with the fluorophores. When heated above the glass transition temperature of the polymer, they can be made to flow. The plastic can then be injected into a mould, or extruded through a nozzle~\cite{yoshimura_1998,thevenin_1980}. This produces lower-quality scintillators, but at much lower cost. As such, these methods are commonly used in the production of large-scale detectors.

These traditional methods have several issues. They require expensive industrial equipment to heat and cast, mould or extrude the plastic. The shape of the scintillator cannot easily be changed. These methods tend to produce blocks or cylinders of a fixed size, which must then be worked to the required shape using subtractive methods like cutting or drilling. The composition of a scintillator is fixed by the manufacturer and cannot be easily changed to suit a particular application. 

\subsection{Additive Manufacturing of Scintillators}
\label{sec:3D-printing}

Additive manufacturing, or 3D-printing, offers a way to solve the problems outlined in Section~\ref{sec:scint_prod}. It allows users to cheaply and rapidly produce scintillators in a scalable manner, and enables easy customization of the composition, internal geometry and external geometry of the scintillator. This is particularly true for plastic scintillators, which can be produced using off-the-shelf equipment. Two main methods have previously been used to 3D-print scintillators: stereolithography (SLA) and fused deposition modelling (FDM)~\cite{zheng_2012,fdm}. Both work on the basic principle of building up a piece, layer by layer. In SLA, an ultraviolet (UV) laser is used to selectively cure a liquid resin. This method allows for easy inclusion of additives such as dyes, fluorophores or neutron-sensitive isotopes. It also offers a high printing resolution, with the smallest feature possible being set by the size of the UV laser spot. However, it can only be used with a limited subset of plastics; the resins that can be used also pose a significant health hazard. SLA printing is also limited in its ability to produce large parts, due to the limited print volumes of available SLA printers.

The simplicity with which additives can be integrated into SLA printing methods has made it a focal area for research into 3D-printed scintillators. Mishnayot et al.~\cite{mishnayot_2014} used a combination of PPO, POPOP and napthalene in a curable resin to obtain an effective printable scintillator. More recently, Kim et al.~\cite{kim_2023} have successfully printed PVT-based scintillators using SLA, obtaining very good results. Stowell et al.~\cite{stowell_2020} have used SLA techniques to manufacture a neutron-sensitive scintillator, using boron nitride and zinc sulfide in an inert matrix. 

FDM, meanwhile, extrudes a thermoplastic filament through a heated nozzle. FDM printing can be used with a wider variety of plastics than SLA, enabling the direct printing of scintillators without additives. It also allows multiple materials or filaments to be printed together. FDM printers are also available at lower cost. FDM is also much more suitable for printing larger parts, as FDM is an easily scalable technology; consumer FDM printers can reach volumes of greater than 600,000~cm$^3$~\cite{elegoo}. The production of customized filaments  requires a separate system; as a result, its use for scintillators has been explored less thoroughly than the use of SLA. Early experiments by Hamel and Lebouteiller~\cite{hamel_2020} used poly-lactic acid doped with PPO and 1,4-Bis (2-methylstyryl) benzene (bis-MSB) to print scintillators, to moderate results. More successfully, Berns et al.~\cite{Berns_2020,Berns_2022} have demonstrated the ability to print a polystyrene-based scintillator with FDM, while Sibilieva et al.~\cite{Sibilieva_2023} have used FDM to print inorganic scintillators in a plastic matrix. The work of Berns et al. has recently been extended by Weber et al.~\cite{Weber2025}, using FDM to produce a segmented detector for particle tracking. This detector was characterized in beam tests by Li et al.~\cite{li2025}, showing comparable light output to similar detectors produced from cast plastic scintillators.

\section{Neutron Detection}
\label{sec:neutrons}

Neutrons are difficult to detect directly. Instead, indirect methods have to be used. One option is to use activation foils, made from isotopes that will be transmuted to a radioactive isotope when they absorb a neutron~\cite{chiesa_2015}. By measuring the radiation output from these foils post-exposure to a neutron source, the neutron flux from the source can be inferred. This is an effective, and easy-to-implement method, but does not allow for real-time measurements of the source flux. A more real-time method is to measure the energy deposited by protons recoiling from neutron impact within the material~\cite{brooks_1959}. This requires the use of a material with a large proportion of hydrogen, making it suitable for use with plastic scintillators. However, the proton signal can be hard to distinguish from gamma rays and other forms of radiation, requiring the use of pulse-shape discrimination techniques to detect it. PSD is a highly effective technique, but is not possible with all plastic scintillators, requiring high concentrations of fluorophores~\cite{hamel2021}. 

The most flexible option is to dope a detector with isotopes like $^3{He}$, $^{10}\mathrm{B}$ or $^6\mathrm{Li}$, which have a high neutron interaction cross-section~\cite{breukers_2013}. Neutron reactions with these nuclei will release charged particles which can then be detected by conventional means, such as scintillators. For this experiment, we will focus on $^6\mathrm{Li}$, which has a typical abundance of 7.6\% in natural lithium. A neutron reaction with $^6\mathrm{Li}$ will release a tritium nucleus and an alpha particle, according to the following equation.

\begin{equation}
\label{eq:x}
\begin{aligned}
^6\mathrm{Li} + \mathrm{n} \rightarrow ^3\mathrm{H} + \alpha.
\end{aligned}
\end{equation}

This reaction produces 4.78~MeV of energy, which is divided between the two reaction products, with $E_{^3H}=2.73$~MeV and $E_\alpha=2.05$~MeV~\cite{zaitseva_2013}. This energy is considerably higher than that from $^{10}\mathrm{B}$ reactions, though quenching within the scintillator will reduce the light output from these significantly~\cite{hamel2021}. It also produces an alpha particle rather than a gamma-ray, meaning that the scintillator light will be emitted close to the interaction. For thermal neutrons, the cross-section of this reaction is 940~barns, giving a strong sensitivity to thermal neutrons. Lithium compounds can easily be integrated into a scintillator, making this interaction a highly suitable method for producing a neutron-sensitive plastic scintillator.

\section{3D-Printing Neutron-Sensitive Scintillators}
\label{sec:3dprinting_neutrons}

We selected FDM for this project, as it is compatible with a wide range of materials. A number of FDM-compatible plastics are effective scintillators, including polyethylene terephalate. However, following the example of Berns et. al (2020)~\cite{Berns_2020}, polystyrene was chosen as the base material for the filament. Polystyrene is well suited to FDM as it has a highly suitable glass transition temperature of $\sim100^\circ $~C and a high melt flow index at FDM temperatures~\cite{aldrich}. It is well understood as a plastic scintillator, and is well-suited to particle detection, having high transparency, short decay times and a high light output for a plastic scintillator. 

\subsection{Filament Production}
\label{sec:filament}

 The polystyrene was combined with PTP as a primary fluorophore and POPOP as a secondary fluorophore. The total photons produced by each scintillator were counted using the detector system described in Section~\ref{sec:detector}, with maximum light output being found to be 30~$\pm$~5~photons/MeV with 2\% of PTP and 0.05\% of POPOP. A pure polystyrene filament proved to be too brittle to effectively 3D-print. It would snap when exposed to the rapid movements of the print head, or be crushed in the extruder, leading to print jams. To increase the flexibility of the filament, a plasticizer was added. Biphenyl, which dissolves well in polystyrene, was chosen. The minimum necessary biphenyl content was found to be 7.5\%, which gave the flexibility required to print effectively without significant impact on light output. Finally, to allow the scintillator to detect neutrons, 0.1\% by weight of $\mathrm{^6Li}$ was added to the material, in the form of $\mathrm{^6LiF}$ powder. All materials were procured from Sigma Aldrich~\cite{aldrichhomepage}.

\begin{figure}
    \centering
    \includegraphics[width=0.3\linewidth]{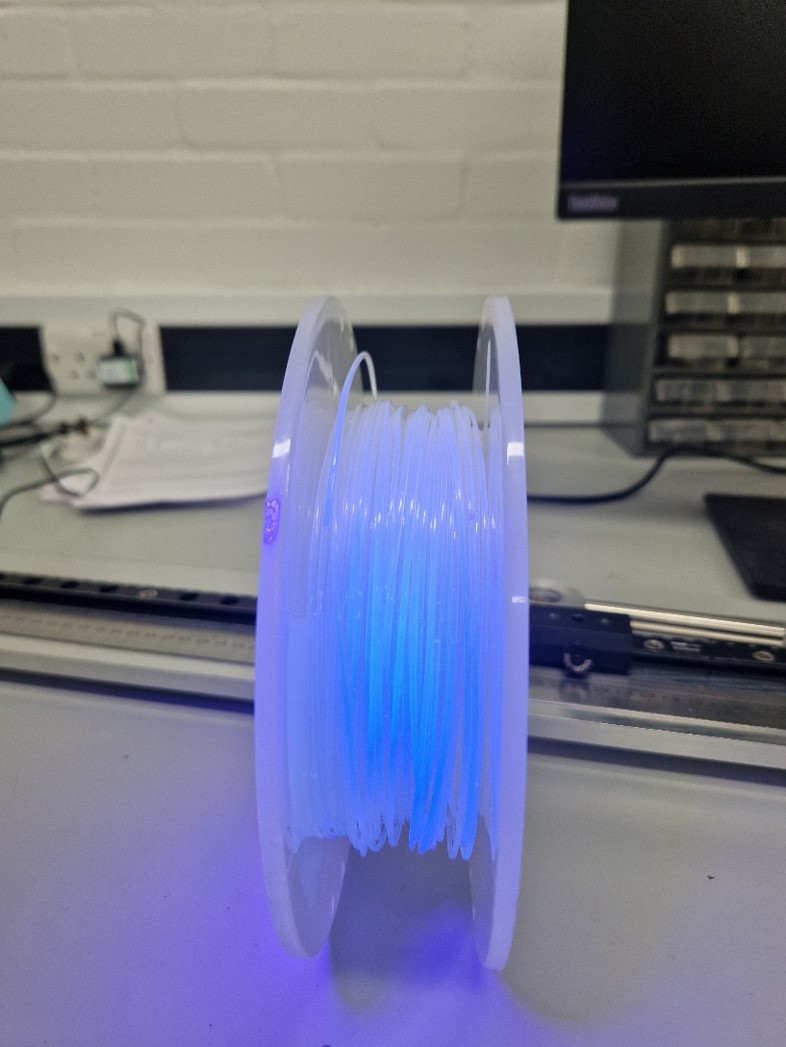}
    \caption{A reel of lithium-containing scintillating filament, fluorescing under UV light}
    \label{fig:filament}
\end{figure}

These components were added to the intake hopper of a 3devo Composer filament maker ~\cite{3devo}, which mixed them and extruded the filament. The filament was extruded in 100~g reels, which could then be printed. An example of one of these reels can be seen in Figure~\ref{fig:filament}. The filament maker, meanwhile, can be seen in Figure~\ref{fig:setup1}. The intake hopper required constant agitation, both to effectively mix the powdered components with the polystyrene pellets and to prevent components with a low melting point from forming a plug which would jam the intake. The filament was extruded using an average temperature of 230$^\circ$~C, at a low speed to ensure adequate cooling of the filament as it was extruded. The 3devo Composer controls the diameter of the filament by controlling the rate at which the filament is drawn through a broad nozzle; this makes it easier to extrude filaments for multiple 3D-printers, which require different filament diameters. For this experiment, only a 1.75~mm diameter filament was produced. 

\subsection{3D-Printing Process}
\label{sec:printing_process}

\begin{figure*}[t!]
    \centering
    \begin{subfigure}{0.5\textwidth}
        \centering
			\centering
			\includegraphics[width=\textwidth, angle=270]{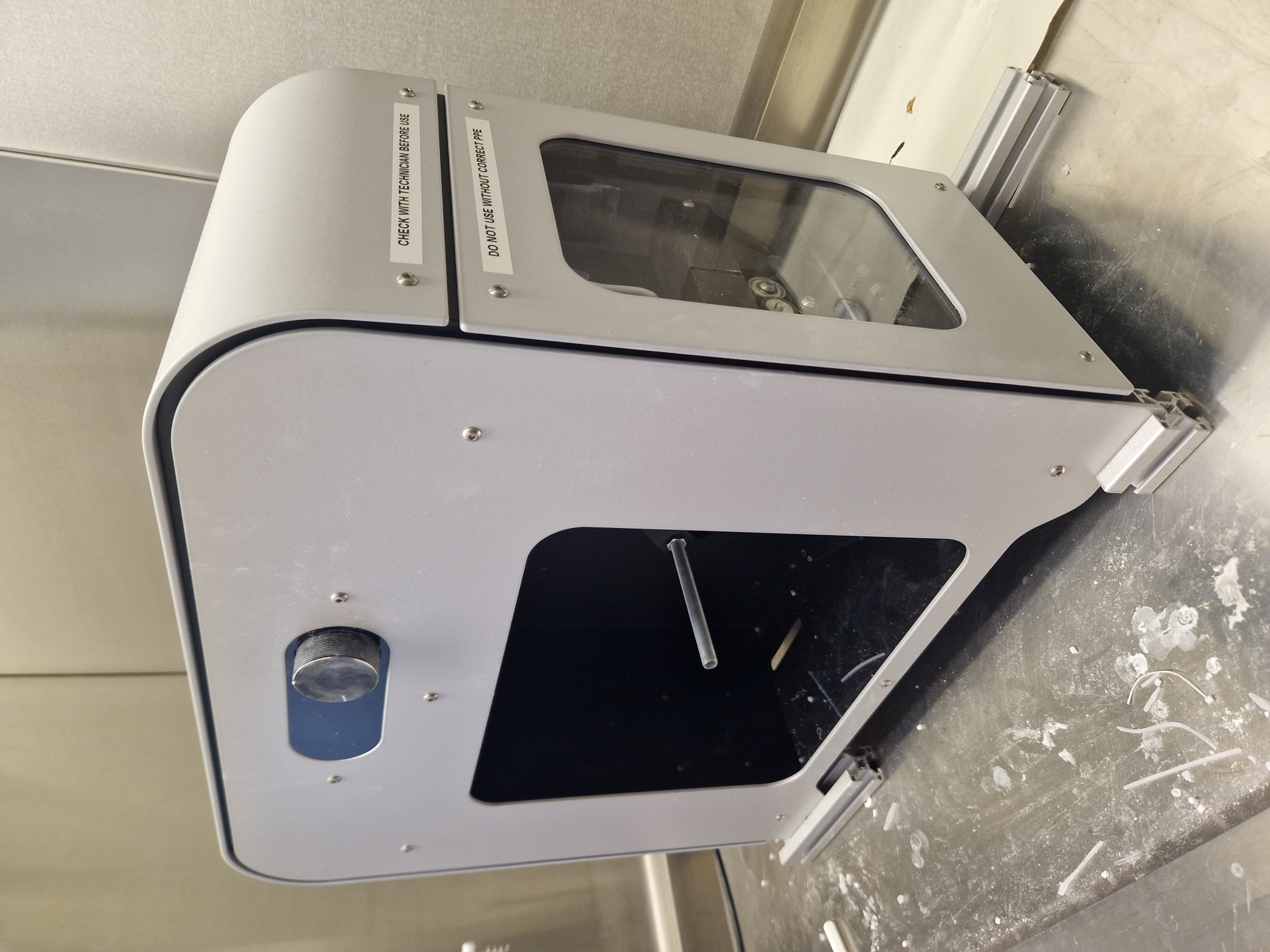}
			\caption{The 3Devo Composer Filament Maker.}
			\label{fig:setup1}
		\end{subfigure}
		\begin{subfigure}{0.34\textwidth}
			\centering
			\includegraphics[width=\textwidth]{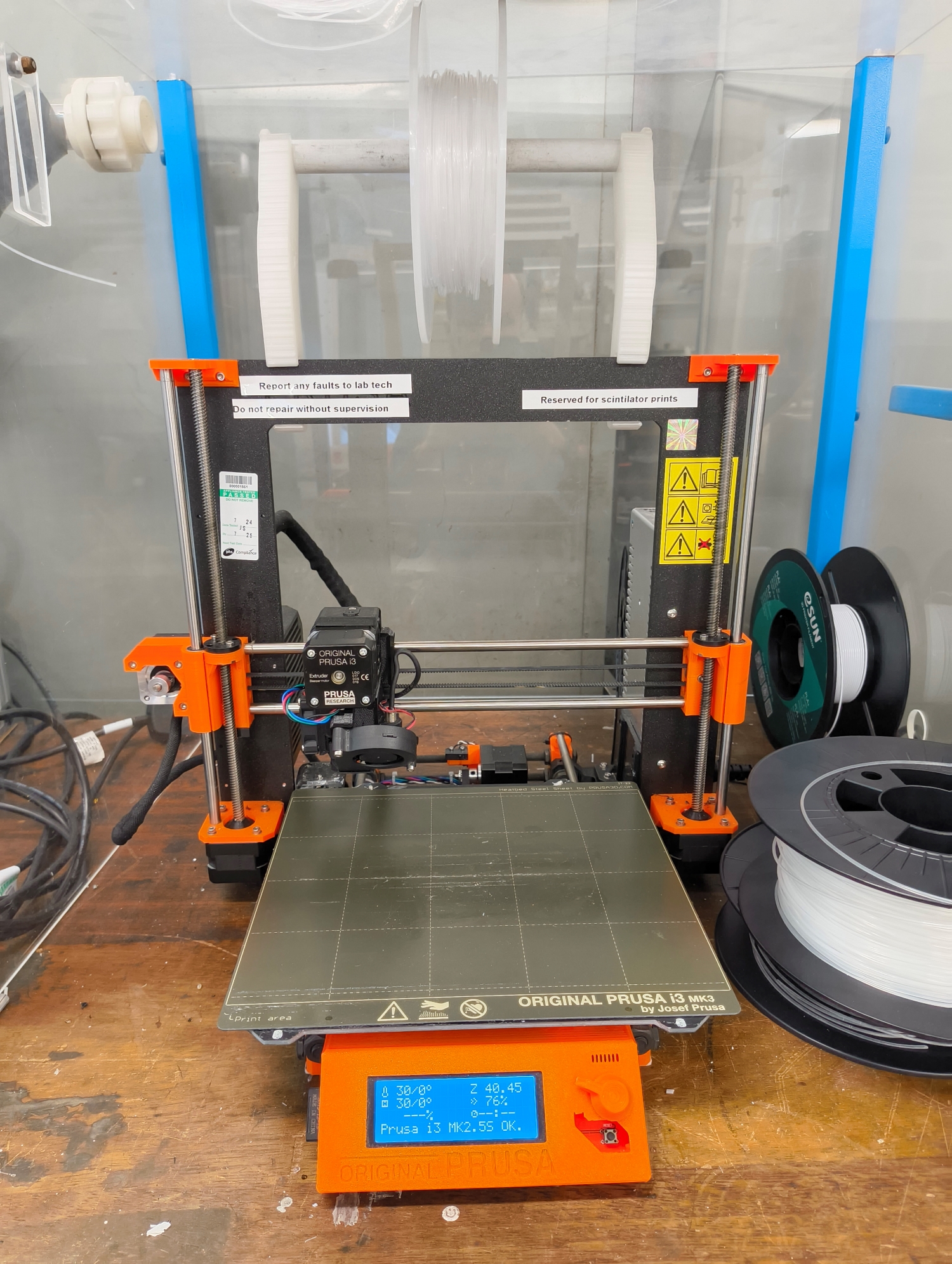}
			\caption{The Prusa i3 3D-printer.}
			\label{fig:setup2}
		\end{subfigure}
		\caption{Figure~\ref{fig:setup1} shows the filament maker used to produce neutron-sensitive 3D-printer filament, with a diameter of 1.75~mm. The filament is then used to print scintillators using the 3D-printer seen in Figure~\ref{fig:setup2}}
		\label{fig:setup}
	\end{figure*}

The filament obtained, as described in Section~\ref{sec:filament}, was printed using a Prusa i3 3D-printer~\cite{Prusa}, as seen in Figure~\ref{fig:setup2}. Most 3D-printed objects use a shell around an infill structure. However, this is unsuitable for printing transparent objects like scintillators, as the infill structure traps gas bubbles. To avoid this, the printing parameters had to be modified to ensure maximum transparency. 100\% infill was selected, to ensure that the final scintillator was filled completely with plastic. To reduce surface roughness, especially on the sides, a layer height of 0.05~mm was selected. The infill layers were also directly aligned, to minimise gas bubbles. There is a strong relationship between printing temperature and transparency, with higher temperatures producing a more transparent scintillator. However, higher temperatures also reduce the shape accuracy of the printing, as the plastic can more easily flow before cooling. A print temperature of 255$^\circ $~C, with no cooling of the print, was found to give the highest transparency while retaining acceptable shape tolerances. To ensure the extruded plastic had time to fill all space on the layer below, the print speed was set to 12~mms$^{-1}$, 20\% of the basic setting. The flow rate was also increased, to 112\% of the base flow rate. This combination of settings was found to give an effectively transparent scintillator. 

However, there were imperfections in the printing of some scintillators, caused by the presence of gas bubbles, which in turn resulted from underextrusion of the plastic. This was due to inconsistencies in the diameter of the filament. These gas bubbles were the main defect noticed during the printing process, and their incidence depended heavily on the quality of the filament batch used for printing, occurring in between 10-25\% of the printed scintillators. Using a filament extruder with a fixed 1.75~mm nozzle, would avoid this issue. Otherwise, it proved to be a reliable method for producing scintillators.

\begin{figure}
    \centering
    \includegraphics[width=0.4\linewidth]{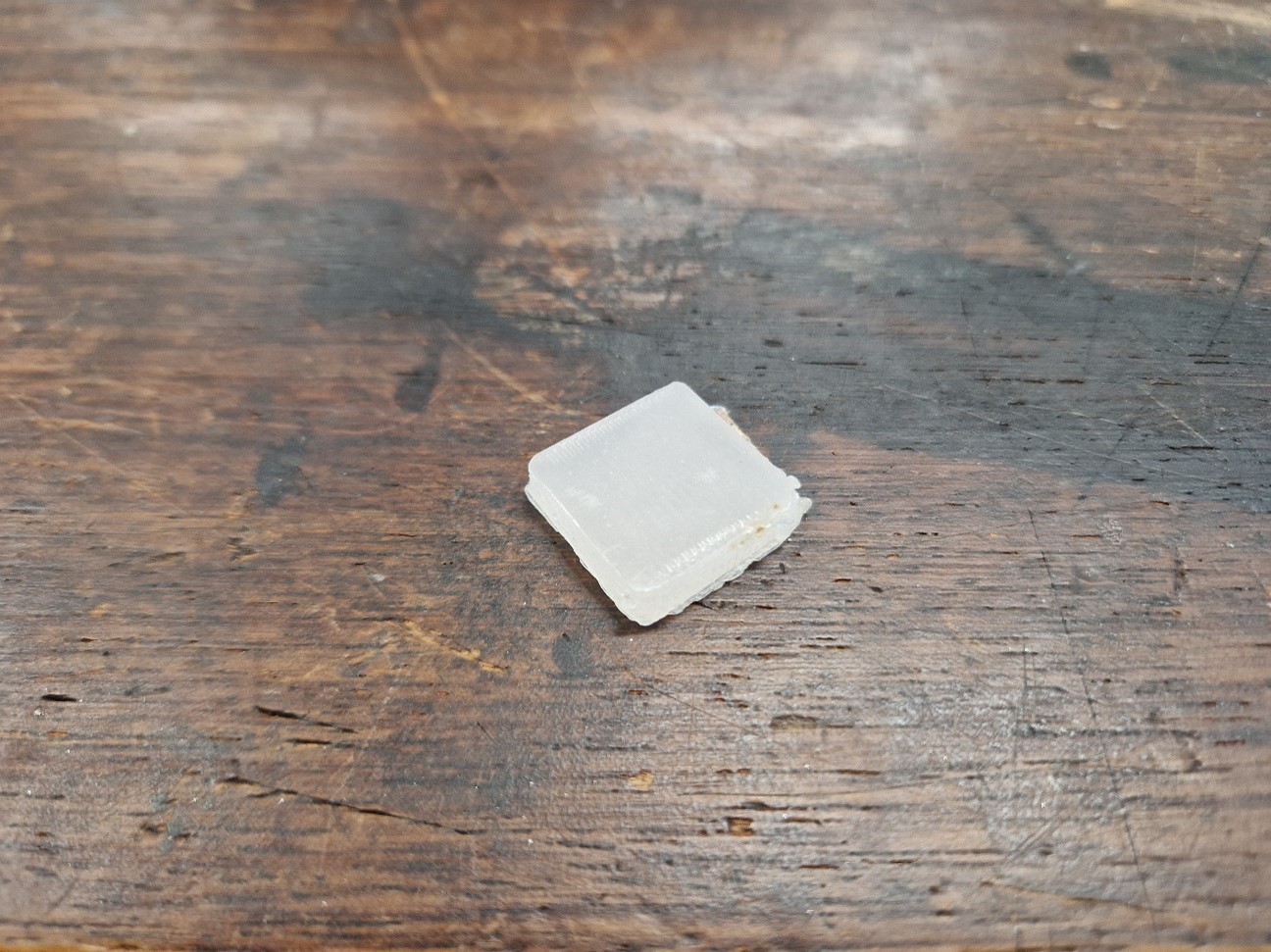}
    \caption{An example of a lithium-containing 3D-printed scintillator; in this case, a 2~cm by 2~cm by 0.5~cm cuboid.}
    \label{fig:scintillator_example}
\end{figure}

After the completion of the 3D-printing process, the scintillator required post-processing to remove roughness caused by the FDM printing process on the external surfaces of the finished part. Additionally, the high printing temperature and high flow rates leave excess plastic on the edges of the printed scintillator. To ensure the print was fully transparent, any excess material was trimmed off, before sanding and polishing. While this reduces the size of the finished scintillator, the polishing process greatly enhances transparency. Accounting for the post-processing requirement, as well as the lack of accurate shape control that results from the high-transparency printing process, the scintillators could be printed to a tolerance of 0.3~mm. A printed and polished scintillator can be seen in Figure~\ref{fig:scintillator_example}. An alternate method to smooth the surface of the scintillator, by exposing it to acetone vapour, was tested. This produced smooth, transparent scintillators that maintained a higher shape tolerance. However, over several days they became cloudy, as residual acetone vapour infiltrated into the interior of the scintillator. Vapour smoothing was also inconsistent; on several pieces, a layer of bubbles formed on the exposed surfaces of the scintillator, resulting in low transparency. 

Scintillators were printed in a variety of shapes and thicknesses. While all displayed sensitivity to $\alpha, \beta$ and $\gamma$ radiation, low sensitivity to neutrons was observed. Simulations indicated that this was due to low lithium content. However, increasing the lithium content would result in decreased transparency, reducing effective light output for a single-material scintillator. This was achieved by printing a two-part scintillator, with a cavity in the centre. This cavity could be packed with $^6\mathrm{LiF}$ powder, increasing the lithium content of the scintillator, but retaining high transparency. The chosen design used a 2~cm by 2~cm by 1~cm cuboid, with a 1~cm by 1~cm by 0.1~cm cavity 0.5~cm inside the scintillator. 

\subsection{Detector Integration}
\label{sec:detector}

\begin{figure}[t]
\centerline{\includegraphics[width = 4in]{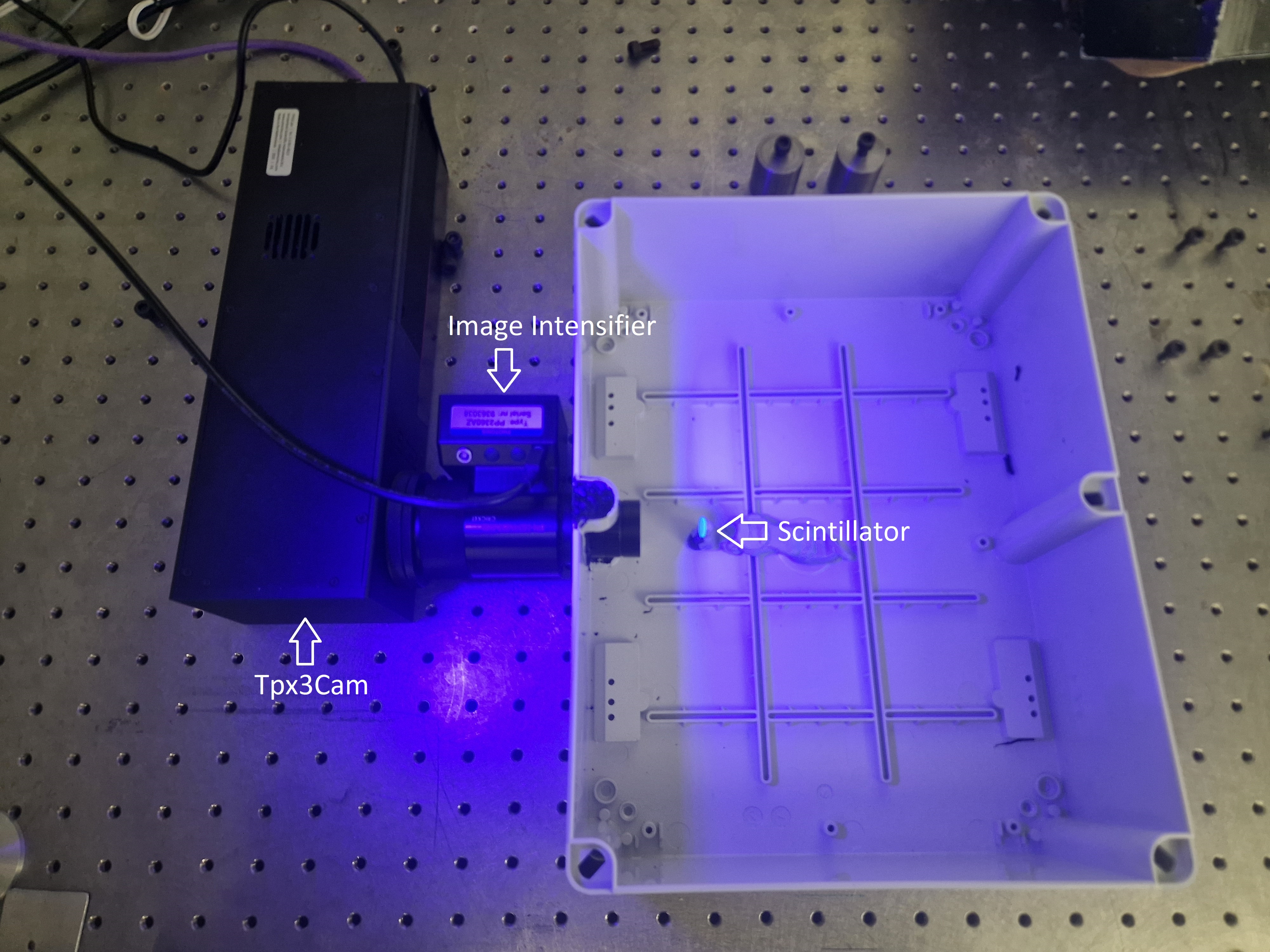}}
\caption{The detector system, illuminated with UV light to show the position of the scintillator. The TPX3Cam is on the left, with the image intensifier connected to it. The scintillator is within the light-tight box, glowing under the UV light.}
\label{fig1}
\end{figure}

Once a satisfactory scintillator was obtained, it could be integrated into a detector system based on an Amsterdam Scientific Instruments TPX3Cam, a fast optical camera based on the TimePix3 (TPX3) readout chip~\cite{timepix}. TimePix3 is an application specific integrated circuit (ASIC) which enables rapid readout of data at high bandwidth, it is bonded to a 256 $\times$ 256 pixel high quantum-efficiency optical sensor~\cite{damen2020}. The spatial resolution of the camera is $\sim$16 $\mu$m and the temporal resolution is 1.56 ns~\cite{Nomerotski_2023}. The TPX3Cam was coupled to a Photonis Cricket\texttrademark ~image intensifier, allowing for single-photon detection~\cite{cricket}. The Cricket\texttrademark ~offers a quantum efficiency of 30\% and a typical gain of 800,000. The gain of the image intensifier is set using a control voltage, which can be varied between 0--5~V; the minimal background noise in this application being found to be at a voltage of 135~mV. To minimise noise due to light contamination, the sensor was placed within a light-tight box, in a position where it could be imaged by the camera, as can be seen in Figure~\ref{fig1}. Initially, a 6~mm wide-angle lens was used to observe the scintillator. This was effective, but was replaced with a macro lens composed of two identical lenses (equivalent aperture f/2) placed back-to-back~\cite{gao2024feasibilitystudynovelthermal}. This led to a significant increase in light collection from the scintillator. 

\section{Detector Performance}
\label{sec:performance}

The scintillator was exposed to radioactive sources including $^{90}$Sr, $^{137}$Cs and $^{252}$Cf, allowing for tests of the its response to $\beta, \gamma$ and neutrons. The scintillating block was placed 10~cm from the sources, in line with the camera lens. The $^{90}$Sr and $^{137}$Cs source were used with no additional items in the light-tight box. The $^{137}$Cs source was used to measure the light yield of the scintillator, counting total photon emission from the scintillator over a fixed exposure time. The $^{252}$Cf source, meanwhile, was contained in a 5~cm thick lead jacket, with a 10~cm High-Density Poly-Ethylene (HDPE) thermalizer between it and the scintillator. A 10~cm lead block was also used to reduce the numbers of gamma rays reaching the scintillator. When a combination of $\gamma$ and neutrons was required, the $^{137}$Cs source was also placed behind the thermalizer, alongside the $^{252}$Cf source. These configurations were simulated using the Geant4 Monte Carlo package, which showed that the thermalizer and lead jacket would create a neutron spectrum with two components; a fast component with energies at around 1~MeV and a slow component with energies at around 0.03~eV~\cite{agostinelli_2003}. The fast component dominated, with 70\% of the neutrons, but the slow component was still significant, containing 20\% of the neutrons, allowing us to test the scintillator's sensitivity to both the fast and thermal neutron components. These simulations, along with all others carried out for this project, used Geant4 11.2, using the \texttt{FTFP\_BERT\_HP} physics list, with the \texttt{G4RadioactiveDecayPhysics} and \texttt{G4OpticalPhysics} options. 

Furthermore, observations of the $^{90}$Sr source were used to determine the optimal composition of the scintillator, testing concentrations of PTP and POPOP between 0-4\% for the former and 0-0.1\% for the latter. As described above, the maximum light output with minimum opacity was found to be given with a concentration of 2\% PTP and 0.05\% POPOP. The $^{90}$Sr and $^{137}$Cs source were then used to determine the detection performance of the scintillator. It was found that the 3D printed materials had a strong sensitivity to both $\beta$ and $\gamma$. The highest light output was found to be $30 \pm5~\mathrm{photons}/\mathrm{MeV}$. This light output is about 10$^4$ times lower than that of NaI. However, the decay time of the emission is significantly shorter~\cite{knoll_2010}. The light output achieved is comparable to that achieved with undoped 3D-printed scintillators by Berns et al.~\cite{Berns_2022}, but is significantly worse than commercially available scintillators produced using traditional methods; for example, Eljen's EJ-270 reaches 4,800~$\mathrm{photons}/\mathrm{MeV}$~\cite{ej270}. 

\begin{figure*}[th!]
    \centering
    \begin{subfigure}[t]{0.75\textwidth}
			\centering
			\includegraphics[width=\textwidth]{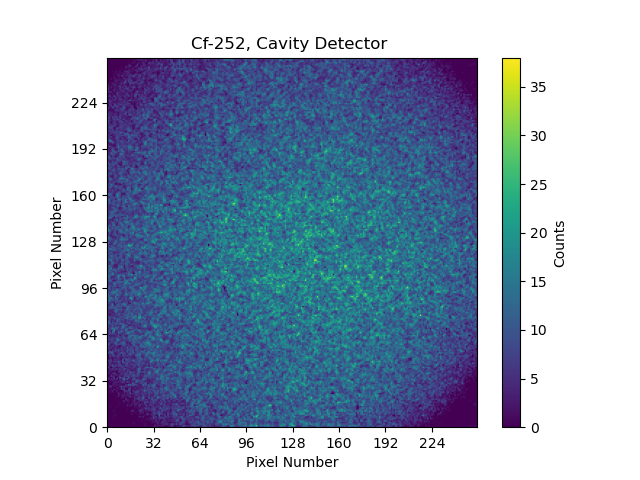}
			\caption{Full image.}
			\label{fig:neutron_image}
		\end{subfigure}
        ~
    \begin{subfigure}[t]{\textwidth}
        \centering
			\centering
			\includegraphics[width=\textwidth]{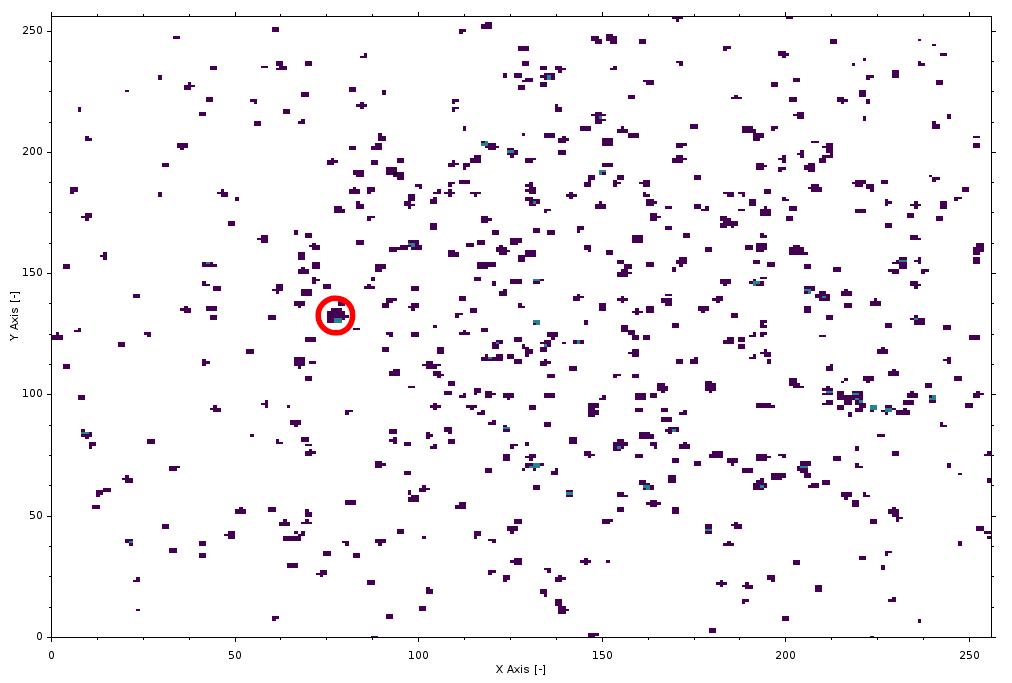}
			\caption{Single frame.}
			\label{fig:single_frame}
		\end{subfigure}
		
		\caption{Two TimePix images of the 3D-printed lithium-packed scintillator, when exposed to a $^{252}$Cf source. Figure~(\ref{fig:neutron_image}) shows a post-processed and summed image, combining multiple such frames from a 30 minute observation. Note the bright central spot, from the scintillator. Figure~(\ref{fig:single_frame}), meanwhile, is a single raw output frame, using a 0.5~s integration time, and shows a single neutron-associated photon cluster to the left of the centre, in the red circle. Colours represent photon counts, with purple being a single photon detected in that pixel and blue two photons.}
		\label{fig:neutron_images}
	\end{figure*}

Measurements of $^{252}$Cf were carried out with two different configurations of the printed scintillator: a 2~cm~$\times$~2~cm~$\times$~0.5~cm cuboid (as seen in Figure~\ref{fig:scintillator_example}, and one with a cavity packed with $^6$Li using the design described in Section~\ref{sec:printing_process}. The characteristic emission of the latter can be seen in Figure~\ref{fig:neutron_image}. These were compared with a commercially produced neutron-sensitive scintillator, Eljen Technology's EJ-420~\cite{ej420}. All three scintillators displayed sensitivity to neutrons, with a significant increase in photo-emission over the noise level. The EJ-420 scintillator, however, was found to be more sensitive than the 3D-printed cuboid, with 4.5 times more detections. The scintillator with the increased Li content was, as expected, more sensitive than the cuboid, increasing detections by a factor of 2. 

% \begin{figure}[t]
% \centerline{\includegraphics[scale=0.5]{images/cf_252_cavity.png}}
% \caption{An image, taken using the TimePix3 camera, showing the 3D-printed lithium-packed scintillator exposed to a $^{252}$Cf source. }
% \label{fig:neutron_image}
% \end{figure}

% \begin{figure}[t]
% \centerline{\includegraphics[scale=0.25]{images/frame_test.png}}
% \caption{A single TimePix frame, imaging the 3D-printed lithium-packed scintillator exposed to a $^{252}$Cf source, showing a single neutron-associated photon cluster to the left of the centre.}
% \label{fig:single_frame}
% \end{figure}

\subsection{Neutron-Gamma Discrimination}
\label{sec:ngd}

The use of the TPX3cam allows a method for neutron-gamma discrimination, without having to tune the scintillator for pulse-shape discrimination. As noted in Section~\ref{sec:neutrons}, a neutron interaction with $^6\mathrm{Li}$ will release more energy than that carried by the typical gamma-ray. As such, when the alpha particle and tritium nucleus are detected within the scintillator, a brighter pulse will be emitted. Given the total reaction energy of 4.78 MeV~\cite{zaitseva_2013} and the scintillator response of $\sim$30 photons per MeV, $\sim$140 photons will be produced for each neutron reaction; though this naive analysis neglects quenching, which will reduce this number significantly. Even so, there will be a bright, compact pulse emitted in the area where the neutron is detected. For gamma rays with a comparable energy to the reaction products, the photons will be emitted over a broader area, resulting in a more diffuse pulse. The compact, bright pulse from the detection of a neutron will be visible as a large cluster in the TimePix output. By identifying the signature of these larger clusters, we can use this to count individual neutron detections. A cluster is defined as a number of hits within a radius of three TimePix pixels within a time resolution of 100~ns; this was chosen as it matches the decay time of the phosphors in the image intensifier. An example of these clusters can be seen in Figure~\ref{fig:single_frame}. By summing the total photon energy for each cluster, we can use this as a proxy for the energy of the incoming particle. In the case of the TimePix, this is in the form of a Time over Threshold (ToT), in nanoseconds.

\begin{figure*}[th!]
    \centering
    \begin{subfigure}[t]{0.8\textwidth}
        \centering
			\centering
			\includegraphics[width=\textwidth]{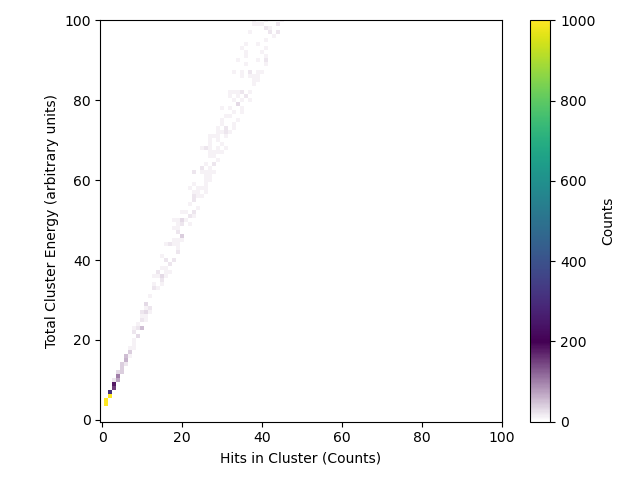}
			\caption{Gamma Rays.}
			\label{fig:total_plot_gamma_sim}
		\end{subfigure}
        ~
		\begin{subfigure}[t]{0.8\textwidth}
			\centering
			\includegraphics[width=\textwidth]{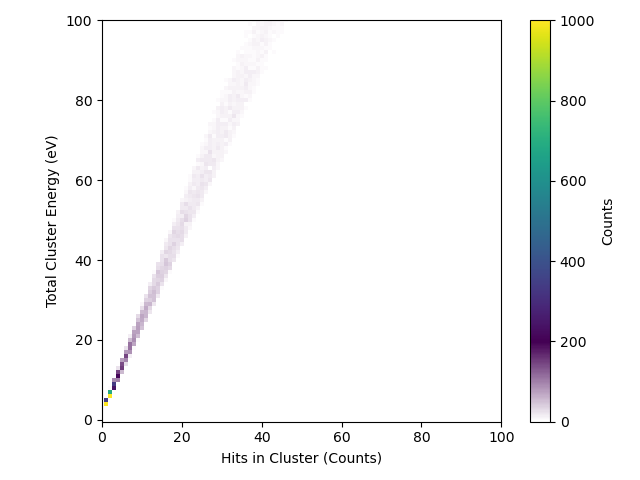}
			\caption{Neutrons.}
			\label{fig:total_plot_neutron_sim}
		\end{subfigure}
		\caption{Representations of cluster size against energy for Geant4 simulated particles interacting with a plastic scintillator. Figure~(\ref{fig:total_plot_gamma_sim}) is for a simulation with gamma rays only, while Figure~(\ref{fig:total_plot_neutron_sim}) is for neutrons only. Both are plotted on the same axes to allow for easy comparison. Note that Figure~(\ref{fig:total_plot_neutron_sim}) has a more extensive 'tail', containing more clusters, above 10 photons per cluster, than is seen with gamma rays, confirming that this can be used to distinguish between the two particles.}
		\label{fig:sim_total_plots}
	\end{figure*}

\begin{figure*}[th!]
    \centering
    \begin{subfigure}[t]{0.8\textwidth}
        \centering
			\centering
			\includegraphics[width=\textwidth]{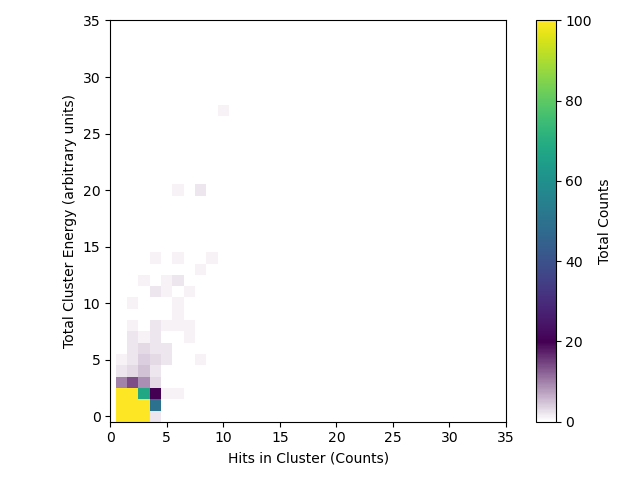}
			\caption{$^{137}$Cs.}
			\label{fig:total_plot_gamma}
		\end{subfigure}
        ~
		\begin{subfigure}[t]{0.8\textwidth}
			\centering
			\includegraphics[width=\textwidth]{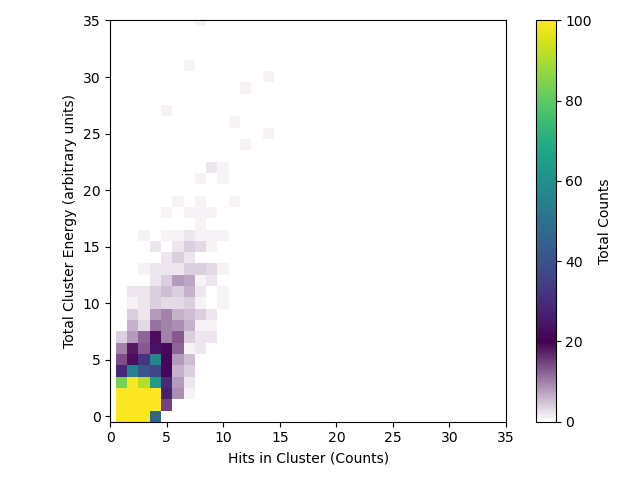}
			\caption{$^{252}$Cf.}
			\label{fig:total_plot_neutron}
		\end{subfigure}
		\caption{Experimental data showing the relationship between cluster size and energy for real particles using the lithium-loaded 3D-printed plastic scintillator. Figure~(\ref{fig:total_plot_gamma}) is for $^{137}$Cs gamma rays only, while Figure~(\ref{fig:total_plot_neutron}) is for $^{252}$Cf and hence includes neutrons. Both are plotted on the same axes to allow for easy comparisons. Note that the brighter 'tail' seen in Figure~(\ref{fig:total_plot_neutron_sim}) is replicated in Figure~(\ref{fig:total_plot_neutron}). Note also that there are significantly more clusters containing 5 or more hits when neutrons are present, making this a useful cut-off for neutron/gamma discrimination. Higher-energy particles create more photons and transfer higher energies, creating the relationship seen here, and allowing for spectral measurements.}
		\label{fig:real_total_plots}
	\end{figure*}

\begin{figure*}[t!]
    \centering
    \begin{subfigure}[t]{0.7\textwidth}
        \centering
			\centering
			\includegraphics[width=\textwidth]{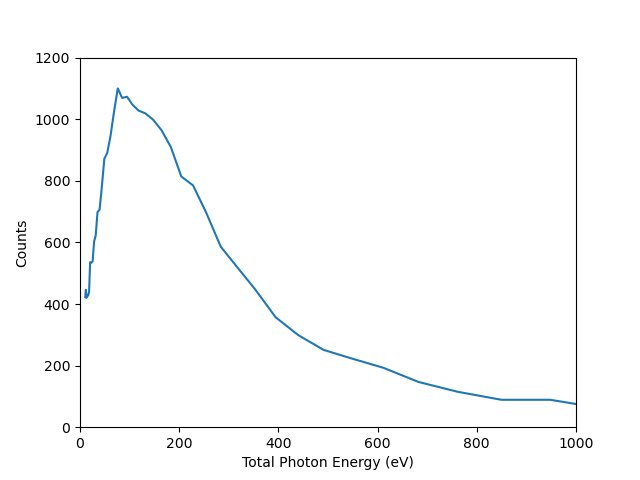}
			\caption{Simulated spectrum.}
			\label{fig:spectrum_sim}
		\end{subfigure}
		\begin{subfigure}[t]{0.7\textwidth}
			\centering
			\includegraphics[width=\textwidth]{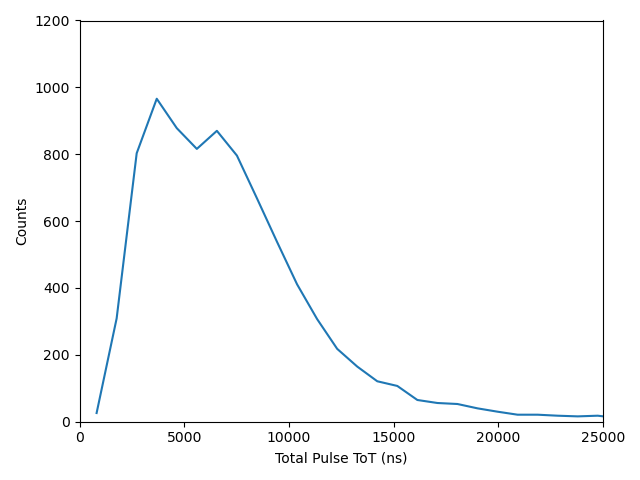}
			\caption{Experimentally measured spectrum.}
			\label{fig:spectrum_real}
		\end{subfigure}
		\caption{Cluster energy spectra, as measured by  from simulation (left,~\ref{fig:spectrum_sim}) and from experimental measurement (right,~\ref{fig:spectrum_real}). Both show a similar peak, which is associated with the presence of thermal neutrons. A double-peak is seen in both figures, but is more pronounced in Figure~\ref{fig:spectrum_real}. This results from light attenuation within the scintillator.}
		\label{fig:spectra}
	\end{figure*}

This was tested with Geant4 simulations of the scintillator, using both thermal neutrons and 662~keV gamma rays. Photons leaving the scintillator and crossing a collector plane were counted, and assigned to clusters, using both time and spatial correlations. The total photon energy was plotted against the total number of collected photons. This showed a clear difference between gamma rays and neutrons, with the neutrons having a 'tail' containing significantly more events, as can be seen in Figure~\ref{fig:sim_total_plots}. This result was also obtained with the 3D printed scintillator, using the $^{137}$Cs and $^{252}$Cf sources to produce gamma-only and neutron/gamma signals, using a 10~cm thick lead block to reduce the number of $^{252}$Cf gamma rays reaching the scintillator. The results of this can be seen in Figure~\ref{fig:real_total_plots}, which shows a clear cut-off in the gamma-ray data at 5 hits per cluster. By setting a minimum value at 5 hits, we are thus able to draw a distinction between neutrons and gamma rays. Based on this, we can plot a cluster energy spectrum for both simulated data and for the real scintillator. Both show a similar peak, associated with thermal neutrons, shown in Figure~\ref{fig:spectra}. $^{252}$Cf fission produces multiple neutrons, which poses only a slight challenge to this method of discrimination. Multiple neutrons arriving simultaneously in widely spaced areas of the scintillator will be detected as separate clusters and counted separately. However, if multiple neutrons arrive in the same location, they will be detected as a single cluster and counted together. This will reduce the number of neutrons counted compared to those detected by the scintillator. 

This method, as it relies on the neutron-$^6$Li reaction, is not sensitive to fast neutrons, as this reaction has only a small cross-section to fast neutrons. Instead, these must be detected using by measuring proton recoil within the scintillator. However, the signature of proton recoil cannot easily be discriminated from gamma ray interactions within the scintillator without changes in composition to ease pulse-shape discrimination using traditional methods.

\clearpage
\clearpage

\clearpage

\section{Conclusions}
\label{sec:conclusions}

This work demonstrates the successful fabrication of a neutron-sensitive scintillating 3D-printed detector via the FDM method, using commercially accessible materials and equipment. The developed scintillator, based on established compounds like Li-6, offers a scalable and rapid production pathway for scintillators. In combination with large commercially available 3D printers, this offers the opportunity to produce scintillators for a wide variety of uses, from small scintillators for handheld detectors to metre-scale scintillators for purposes such as neutrino detectors~\cite{elegoo}. Integration with a fast optical detector system confirmed sensitivity to both ionizing radiation and thermal neutrons, achieving a measured light yield of 30 $\pm$ 5 photons/MeV. Furthermore, a novel neutron/gamma discrimination algorithm was implemented, enabling effective separation of neutron signals from gamma-ray backgrounds, validating the detector’s performance in mixed radiation fields. To improve the performance of the scintillator, we are exploring the use of perovskite materials. This will improve the light output of the scintillator, while preserving the low decay times.

\acknowledgments

This work was funded by a grant from the Nuclear Security Science Network. A. Barr thanks Paul Campbell and Andy McFarlane for support and assistance with printing, materials and sources.

% \paragraph{Note added.} This is also a good position for notes added
% after the paper has been written.

% Bibliography

%% [A] Recommended: using JHEP.bst file
 \bibliographystyle{JHEP}
 \bibliography{3d_printing_bib}

%% or
%% [B] Manual formatting (see below)
%% (i) We suggest to always provide author, title and journal data or doi:
%% in short all the informations that clearly identify a document.
%% (ii) please avoid comments such as "For a review'', "For some examples",
%% "and references therein" or move them in the text. In general, please leave only references in the bibliography and move all
%% accessory text in footnotes.
%% (iii) Also, please have only one work for each \bibitem.

% \begin{thebibliography}{99}

% \bibitem{a}
% Author,
% \emph{Title},
% \emph{J. Abbrev.} {\bf vol} (year) pg.

% \bibitem{b}
% Author,
% \emph{Title},
% arxiv:1234.5678.

% \bibitem{c}
% Author,
% \emph{Title},
% Publisher (year).

% \end{thebibliography}
\end{document}